\documentclass[11pt]{elsarticle}

\pdfoutput=1

\makeatletter
\def\ps@pprintTitle{%
  \let\@oddhead\@empty
  \let\@evenhead\@empty
  \let\@oddfoot\@empty
  \let\@evenfoot\@oddfoot
}
\makeatother

\usepackage{url}
\usepackage{breakurl}
\usepackage[breaklinks,
            colorlinks = true,
            linkcolor = blue,
            urlcolor  = blue,
            citecolor = blue,
            anchorcolor = blue]{hyperref}

\usepackage{lineno}
\usepackage{textcomp}
\usepackage{graphicx}
\usepackage[margin=1.25in]{geometry}
\usepackage[usenames,dvipsnames]{color}
\usepackage{fancyhdr}
\fancypagestyle{plain}{%
  \fancyhf{}%
  \fancyhead[C]{}
  \fancyfoot[C]{\thepage}
}

\fancypagestyle{empty}{%
  \fancyhf{}%
  \fancyhead[C]{{\it Activity report on the African School of Physics, November 28 -- December 9, 2022}}
  \fancyfoot[C]{\thepage}
}
\begin{document}

\begin{frontmatter}

%\pubblock

\title{Activity Report on the Seventh African School of Fundamental Physics and Applications (ASP2022)}

\author[add1]{K\'et\'evi A. Assamagan\corref{cor1}}
\ead{ketevi@bnl.gov}
\author[add2]{Bobby Acharya}
\author[add3]{Kenneth Cecire}
\author[add4]{Christine Darve}
\author[add5]{Fernando Ferroni}
\author[add6]{Julia Ann Gray}
\author[add7]{Azwinndini Muronga}

\cortext[cor1]{Contact}

\address[add1]{Brookhaven National Laboratory, USA}
\address[add2]{ICTP, Italy, and King's College London, UK}
\address[add3]{University of Notre Dame, USA}
\address[add4]{European Spallation Source, Sweden}
\address[add5]{INFN-GSSI, Italy}
\address[add6]{ASP International Advisory Committee, Switzerland}
\address[add7]{Nelson Mandela University, South Africa}

\begin{abstract}
\noindent 
The African School of Fundamental Physics and Applications, also known as the African School of Physics (ASP), was initiated in 2010, as a three-week biennial event, to offer additional training in fundamental and applied physics to African students with a minimum of three-year university education. Since its inception, ASP has grown to be much more than a school. ASP has become a series of activities and events with directed ethos towards physics as an engine for development in Africa. We report on the seven African School of Physics, ASP2022, organized at Nelson Mandela University, on November~28 to December~8, 2022. ASP2022 included programs for university students, high school teachers and high school pupils.
\end{abstract}


\end{frontmatter}

%\linenumbers
%

%\def\thefootnote{\fnsymbol{footnote}}
%\setcounter{footnote}{0}

%\newpage

\section{Introduction}
\label{sec:intro}

The African School of Physics is a collection of activities to support academic growths of African students. One activity is a three-week biennial event organized in different African countries---this event consists of a 2-week intensive school, complemented with a one-week African Conference on Fundamental and Applied Physics (ACP)~\cite{ASP2021-reports, ASP, ASP-reports, asp2018}.  The host country of the next biennial event is selected two and half years in advance through a bidding process. In December 2019, South Africa was selected, from four competing countries, to host the seventh edition of ASP at Nelson Mandela University (NMU) in Gqeberha.  ASP2022 was originally planned in July 2022, for ACP2022 to be held jointly with the South African Institute of Physics annual meeting; however, this plan changed due to travel restrictions and uncertainties resulting from the COVID-19 pandemic: 
\begin{enumerate}
    \item ASP2020, originally planned in Morocco, was cancelled and organized as online event on July 19-30, 2021, and the 3rd week of ASP2021, which would have been ACP2021, was set in December 2021. We decided to split the 3-week term ASP into two distinct events, the school and the conference, held at different times, to avoid the eventuality of having three-long weeks of online engagements, should the COVID-19 pandemic prevent in-person activities; 
    \item ACP2021 was planned as an in-person event, however, the onset of the Omicron variant led to the postponement of ACP2021 to March 2022. Ultimately, uncertainties in travel restrictions forced us to organize ACP2021 as an online event on March 7-11, 2022~\cite{acp2021}; 
    \item With ASP2020/2021 delayed well into 2022, we had to re-consider the dates for the 2022 edition of ASP, which was planned as two events—one, an intensive school, on November 28 to December 9, 2022 (ASP2022), and other, a conference on September 23-29, 2023 (ACP2023). 
\end{enumerate}

In this paper, we present the activity report of ASP2022. In Section~\ref{sec:prog}, we review the scientific program and discuss the supports received in Section~\ref{sec:sup}. We present the profiles of the participants and expenditures in Sections~\ref{sec:prof} and~\ref{sec:exp} respectively. Feedback from participants are presented in Section~\ref{sec:feed}. Outlook and conclusions are offered in Sections~\ref{sec:out} and~\ref{sec:conc}.

\section{Scientific Program}
\label{sec:prog}

As mentioned in Section~\ref{sec:intro}, The African School of Physics has evolved well beyond three-week biennial engagements. For broader participation in fundamental fields and related applications, the scientific program includes the major physics areas of interest in Africa, as defined by the African Physical Society (AfPS)~\cite{AfPS}:
\begin{itemize}
   \item Particles and related applications: nuclear physics, particle physics, medical physics, (particle)astrophysics \& cosmology, fluid \& plasma physics, complex systems;
   \item Light sources and their applications: light sources, condensed matter \& materials physics, atomic \& molecular physics, optics \& photonics, physics of earth;
   \item Cross-cutting fields: accelerator physics, computing, instrumentation \& detectors.
\end{itemize}
Topics in quantum computing \& quantum information and machine learning \& artificial intelligence are also on the agenda. Furthermore, the ASP program includes the fields of societal engagements, namely: topics related to physics education, community engagement, women in physics, early career physicists and engagements with African policymakers in research and education. Representative details on the scientific program are presented in Ref.~\cite{acp2021}, and in the references therein.

The scientific program was arranged in three distinct components for university students, high school teachers and high school pupils. The program for university students was set up for the entire duration of the school, from November~28 to December~9, 2022. The program for high school teachers was in parallel to the students’ program, with some lectures in parallel, during the period of November~28 to December~2, 2022. During the week of December 5--9, 2022, we organized an outreach event for high school pupils, in parallel to the students’ program. On December~3, 2022, we organized a forum where we discussed the sustainability of ASP, and capacity development and retention in Africa. Details of the scientific agenda can be found in Ref.~\cite{ScientificProgram}. 
\subsection{Scientific program for students}
This program was organized as a series of plenary and parallel sessions, hands-on tutorials and experiments, in fundamental and applied physics: particles and related applications, astrophysics and cosmology, condensed matter, materials physics and related applications, accelerators, detectors, simulations and high-performance computing. These areas of fundamental and applied physics were selected, considering their academic concentrations of African students as discussed in Section~\ref{sec:prof}. 
\begin{figure}[!htbp]
 \begin{center}
  \includegraphics[width=\textwidth]{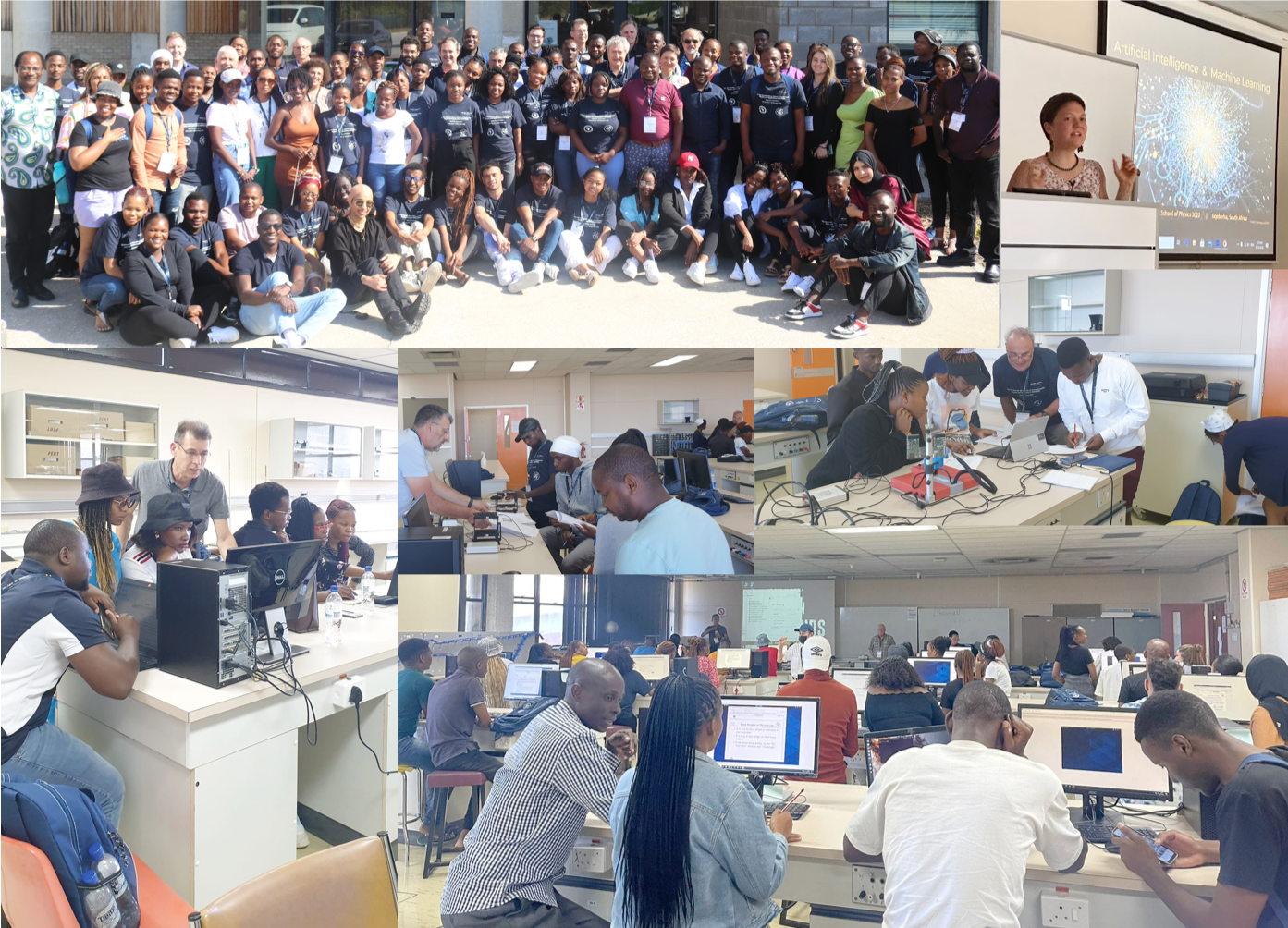}
   \caption{Interactions between students and lecturers during ASP2022.}
    \label{fig:students}
  \end{center}
\end{figure}
To better support the students, topics of general interests were arranged in plenary sessions and advanced topics were addressed in parallel sessions. This organization of the lectures allowed the students to follow in depth topics that support the academic majors; it allowed coverage of a remarkable amount of materials in the short duration of a two-week engagement. Figure~\ref{fig:students} shows snapshots of students' activities.
\subsection{Scientific program for high school teachers}
\begin{figure}[!htbp]
 \begin{center}
  \includegraphics[width=\textwidth]{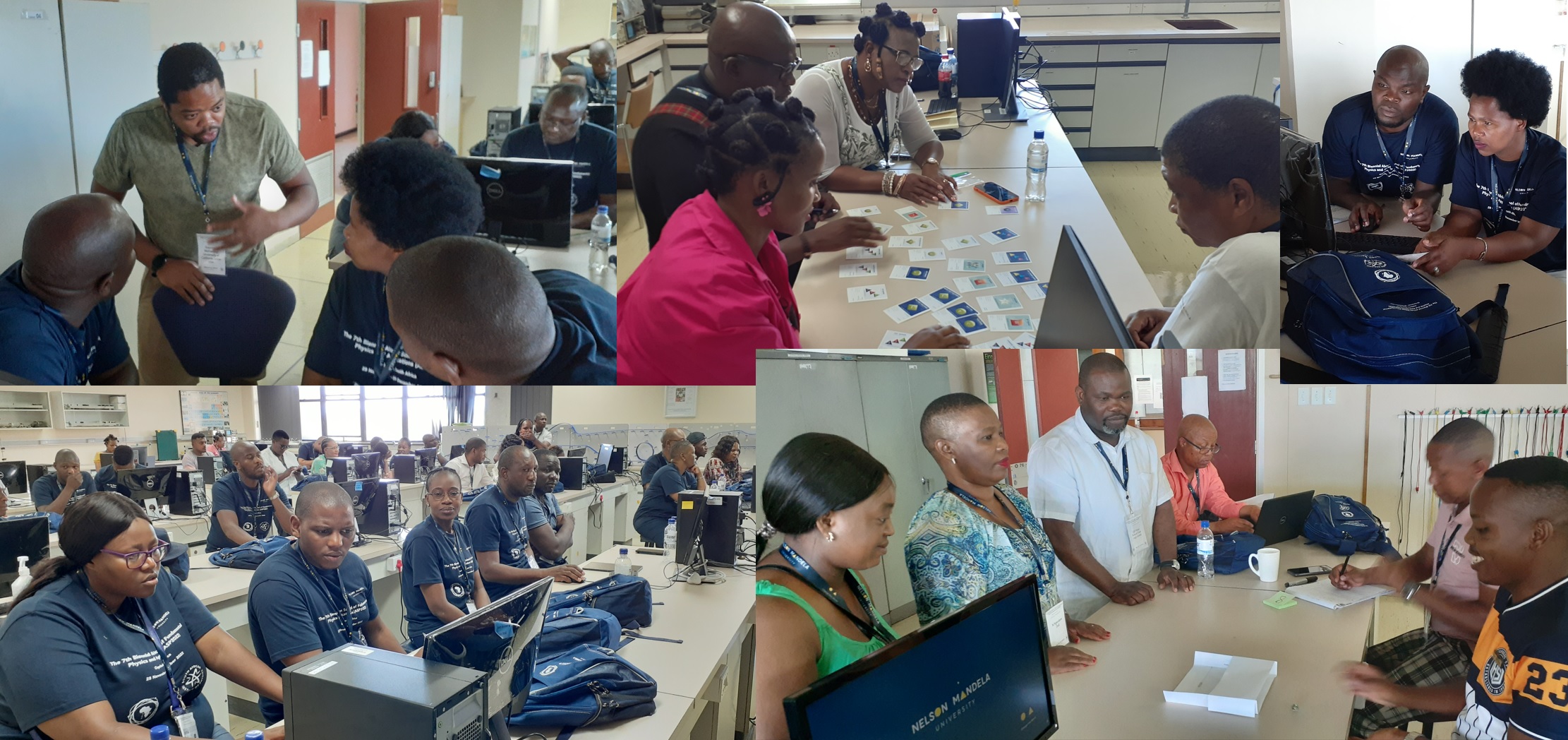}
   \caption{High school teachers during ASP2022.}
    \label{fig:teachers}
  \end{center}
\end{figure}

The objective of the teachers' program is to help in their planning and delivery of physics instruction. Practically, there were two parts to this prgogram. The South African Department of Education the South African Institute of Physics (SAIP) identified the physics topics in which teachers’ development is much needed, based on student results on standard examinations. Two facilitators from the University of Limpopo gave classes covering these topics. The second part focused on contemporary physics to extend teachers' breadth of understanding and to offer new ways to engage students: the very topics students study in high school become the tools to analyze physics results at the cutting edge. Two facilitators from the U.S. QuarkNet program~\cite{QuarkNet} used physics data activities and masterclasses developed for teachers through QuarkNet. Details on the scientific program for high school teachers are in Ref.~\cite{ScientificProgram}. A group of 76 South African teachers, most from Eastern Cape, participated. Figure~\ref{fig:teachers} shows high school teachers during ASP2022.
\subsection{Scientific program for high school pupils}
\begin{figure}[!htb]
 \begin{center}
  \includegraphics[width=\textwidth]{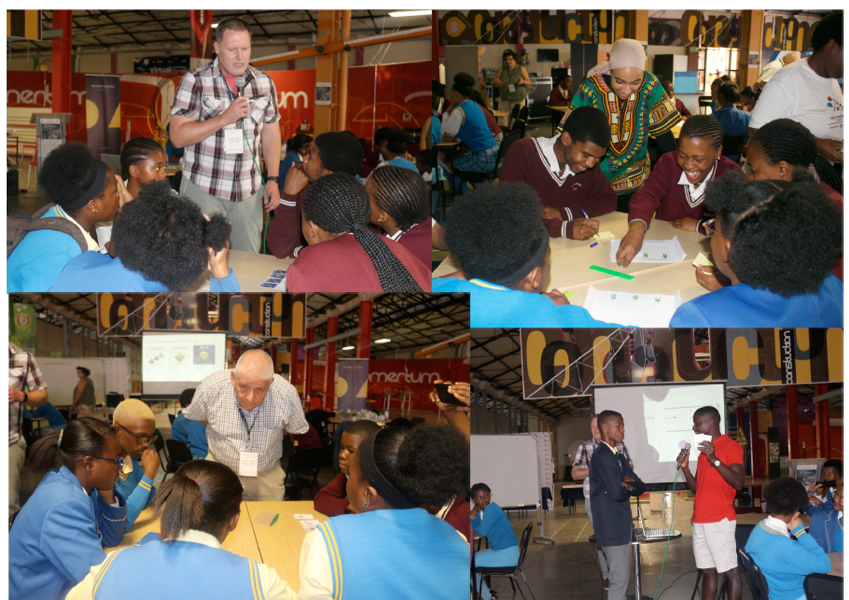}
   \caption{Engagement with high school pupils during ASP2022.}
    \label{fig:learners}
  \end{center}
\end{figure}

The objective for the scientific program for high school pupils is to encourage them to develop and maintain interest in physics. This program was organized as an outreach event with hands-on activities in fundamental physics developed by QuarkNet and information on fundamental physics research based in Africa. The outreach activities, facilitated by the two QuarkNet staff members in the Teachers' Program and a group of ASP2022 Lecturers, included particle cards, cosmic muon experiment, particle-collision activities and questions and answers session ~\cite{ScientificProgram}. The Q\&A was especially motivating for students as they could "ask anything" of physicists who were mostly young and mostly from Africa. The program was repeated for four hours daily for five days for different groups of pupils from schools in and around Gqeberha. A total of 231 pupils attended. Figure~\ref{fig:learners} shows physics outreach sessions with high school learners.
\subsection{ASP Forum}
The objective of the ASP Forum is to engage policymakers in physics education and research during the school. For this particular event, two major topics were discussed at the forum: mechanism to make ASP sustainable in the long term and retention of African physics graduates and faculties. Representatives of policymakers in Morocco, Senegal, Ivory coast, Burkina Faso, Benin and South Africa (DSI, NRF, SAIP, SANSA, NMU), and international delegates from Africa, Europe and the U.S. participated in the discussions. The long-term sustainability of ASP requires not just African participation in ASP events, but also significant financial contributions from African countries, not just the host countries. Such financial contributions need to not be large, but rather consistent from all African countries: it was pointed out that a per country contribution of \$5000 every two years would be enough when added to external contributions. Discussions will continue between the countries represented at the forum. 

\begin{figure}[!htb]
 \begin{center}
  \includegraphics[width=\textwidth]{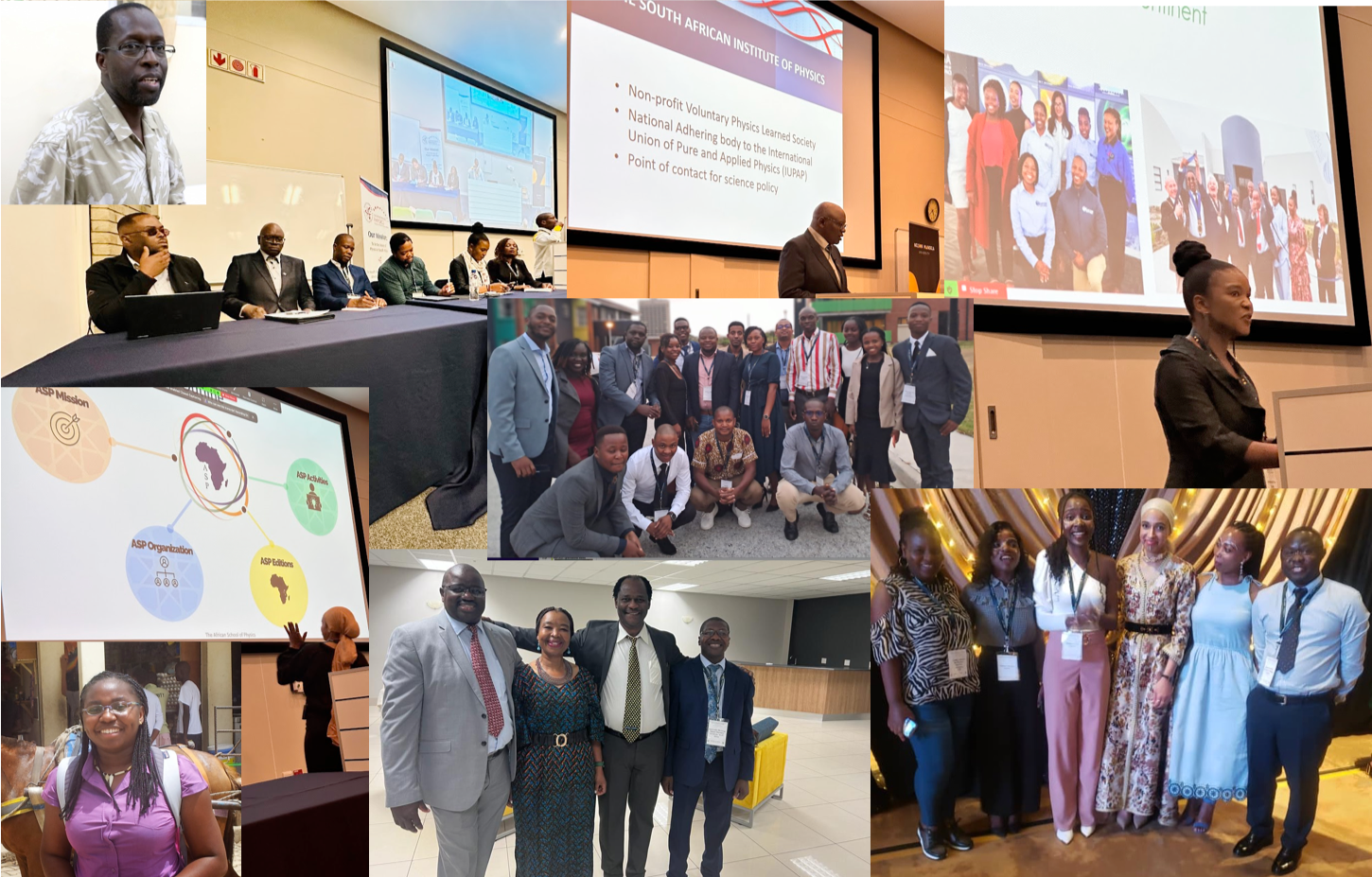}
   \caption{ASP forum for discussion with policymakers on capacity development and retention in Africa.}
    \label{fig:forum}
  \end{center}
\end{figure}

The issue of retention relates to how best to use trained African faculties and researchers in the development of education and research in Africa. Retention does not necessarily mean a return to Africa; rather, African physicists, wherever they are, should find or establish impactful avenues for their contributions to African scientific development.  The agenda of the ASP2022 Forum can be in Ref.~\cite{ScientificProgram}. At the conclusion of the forum, we enjoyed a gala dinner, sponsored and hosted by NMU. Figure~\ref{fig:forum} shows of participants at the ASP2022 forum and dinner on December 3, 2022.
\section{Support}
\label{sec:sup}

\begin{figure}[!htbp]
 \begin{center}
  \includegraphics[width=\textwidth]{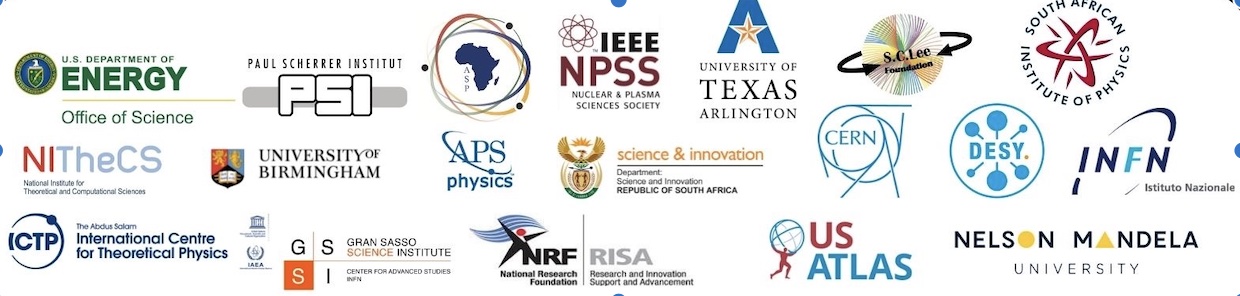}
   \caption{Logos of the institutes that supported ASP2022 financially or in-kind. BNL and FNAL also contributed to the ASP2022 budget.}
    \label{fig:logos}
  \end{center}
\end{figure}

We received financial and in-kind support from various institutes whose logos are shown in Figure~\ref{fig:logos}. In addition, there were contributions from Brookhaven National Laboratory (BNL), Fermi National Laboratory (FNAL), and the MacDonald Institute. The support received allowed full coverage of the expenses associated with the event as presented in Section~\ref{sec:exp}. 
In addition, some lecturers received travel coverage from their institutes.

\section{Participant profiles}
\label{sec:prof}
For the high school outreach program, 231 high school pupils---of the tenth to the twelfth grades, some of whom are shown in Figure~\ref{fig:learners}---participated, with an average of 46 pupils per day during the period of December 5-9, 2022. We had planned to accommodate a much larger number of pupils, but the timing of the event coincided with the end-of-the-year school break and that affected the level of participation. The outreach activities were carried out in the mornings up to lunch time, at two venues, namely Nelson Mandela University, Missionvale Campus (2 days) and Nelson Mandela Bay Science and Technology Centre in Kariega (3 days). Most of the pupils were exposed to these scientific activities for the first time and this broadened their views of physics and science in general.

Originally, we targeted eighty high school teachers; ultimately, seventy-six teachers from South Africa attended the event in person, as shown Figure~\ref{fig:teachers}. These teachers were identified by the education authorities. The teachers were selected from the Eastern Cape region of South Africa to facilitate transport logistics to Gqeberha. 

\begin{figure}[!htbp]
 \begin{center}
  \includegraphics[width=\textwidth]{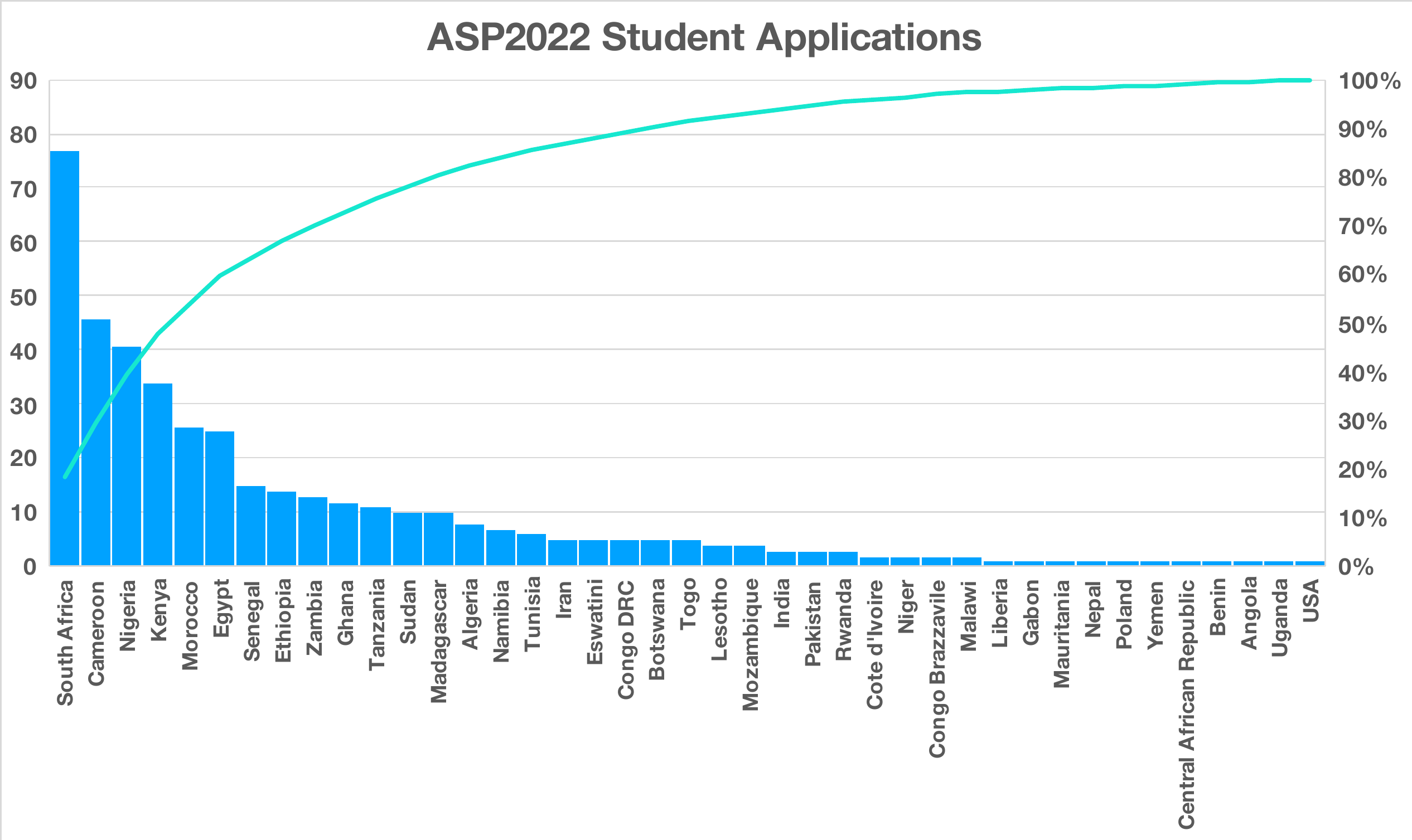}
   \caption{Distribution of ASP2022 student applications as a function of their countries of citizenship.}
    \label{fig:applications}
  \end{center}
\end{figure}

\begin{figure}[!htbp]
 \begin{center}
  \includegraphics[width=\textwidth]{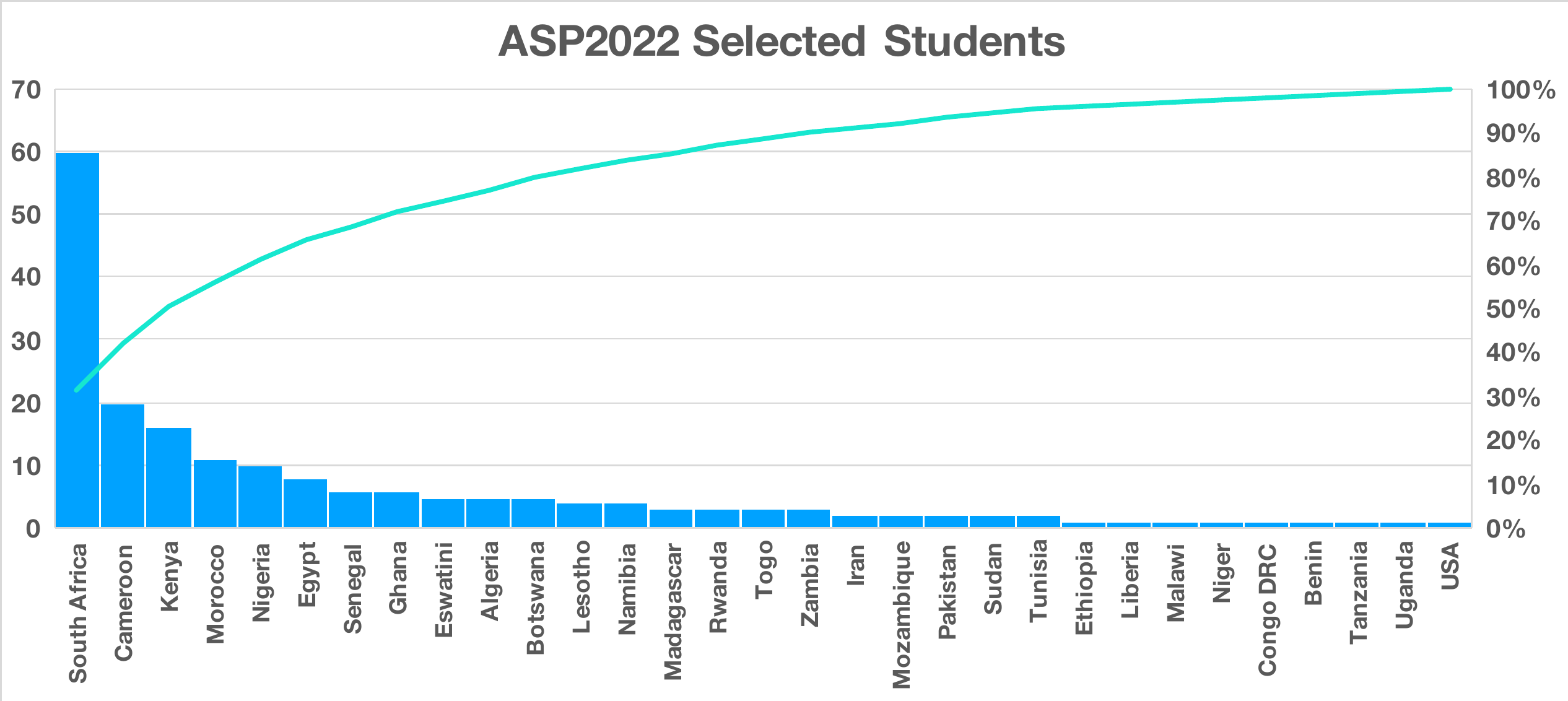}
   \caption{Distribution of selected students for ASP2022 as a function of country of citizenship.}
    \label{fig:selections}
  \end{center}
\end{figure}

\begin{figure}[!htb]
 \begin{center}
  \includegraphics[width=\textwidth]{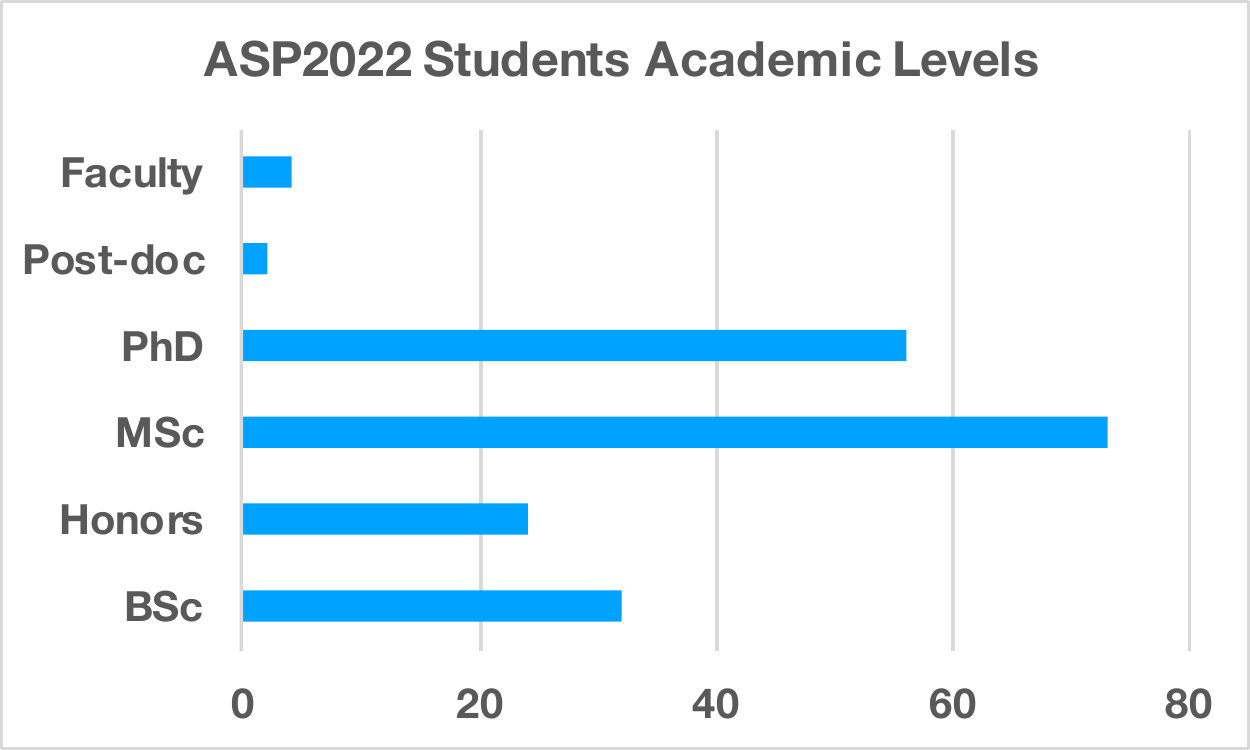}
   \caption{Academic levels of the selected students at ASP2022.}
    \label{fig:degrees}
  \end{center}
\end{figure}

\begin{figure}[!htb]
 \begin{center}
  \includegraphics[width=\textwidth]{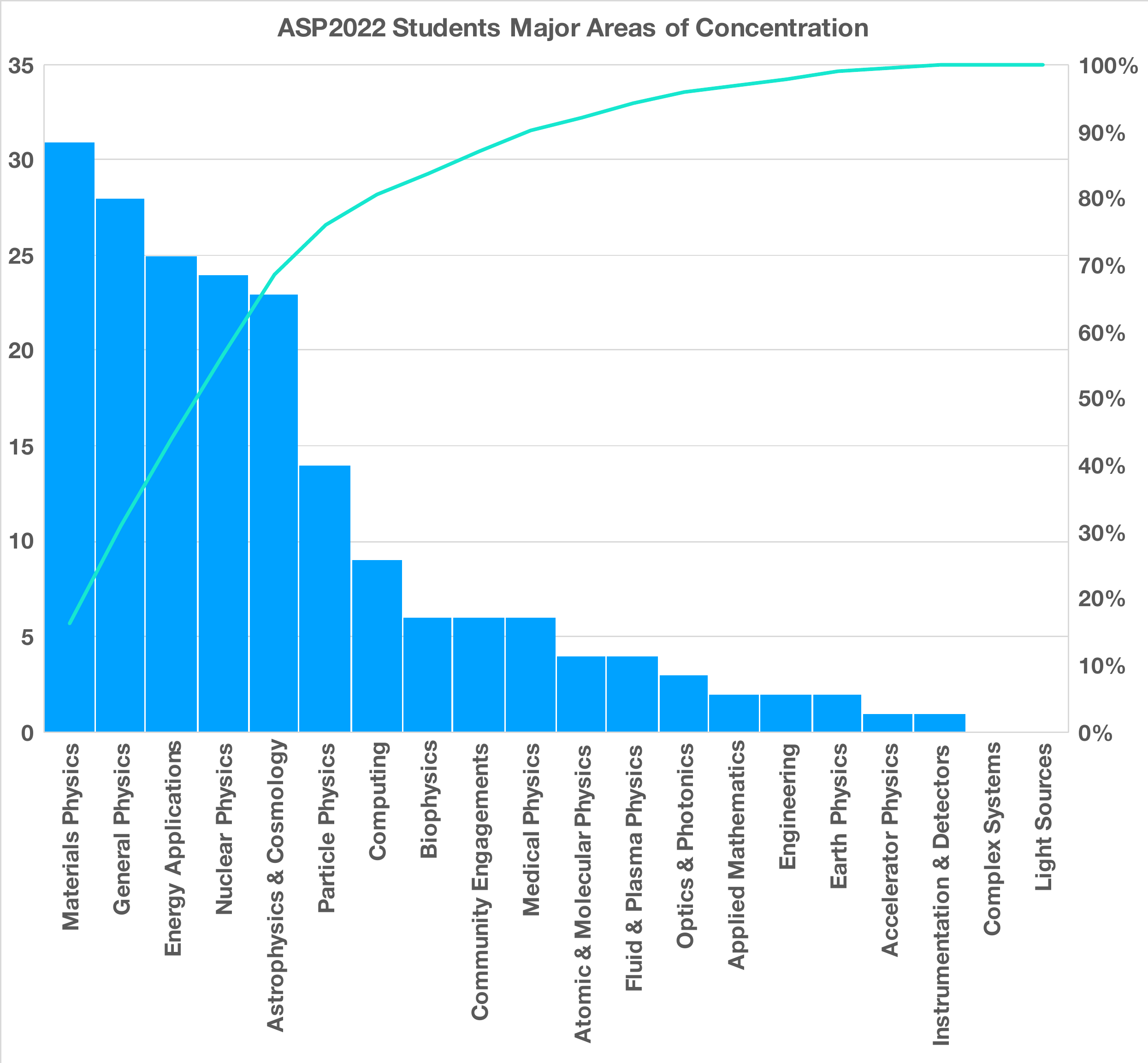}
   \caption{Academic concentrations of the selected students at ASP2022.}
    \label{fig:majors}
  \end{center}
\end{figure}

Upwards of 416 candidates from 41 countries applied for ASP2022 as shown in Figure~\ref{fig:applications}.
All the applications went through a rigorous selection process where they had to submit their curriculum vitae, university transcripts, a letter of motivation and arrange for one letter of recommendation. The selection committee was subdivided into subcommittees assigned to review a subset of the applications according to selection criteria set by the International Organizing Committee (IOC). Members of the selection committee were volunteers from the lecturers and the Local Organizing Committee (LOC). One hundred and ninety-one applicants were selected as shown in Figure~\ref{fig:selections}.
The female-to-male ratio of the selected students was 9:10. Of the 191 selected students, there were just four declinations. A few students (and also lecturers) had issues in obtaining entrance visa for South Africa and could not attend in person. The selected students were required to have a minimum of three-year university education in engineering, computing, and fundamental and applied physics. Backgrounds on the selected students are shown in Figures~\ref{fig:degrees} and~\ref{fig:majors}.

\section{Expenditures}
\label{sec:exp}
The financial support received and described in Section~\ref{sec:sup} was used to over expenses for ASP2022. These include travels and full room and board for in-person students, internet connectivity for online participant, online networking, travels and/or lodging accommodation for some lecturers, local transportation, small detector lab equipment for students, teachers, and pupils. 

\section{Feedback}
\label{sec:feed}
\begin{figure}[!htbp]
 \begin{center}
  \includegraphics[width=\textwidth]{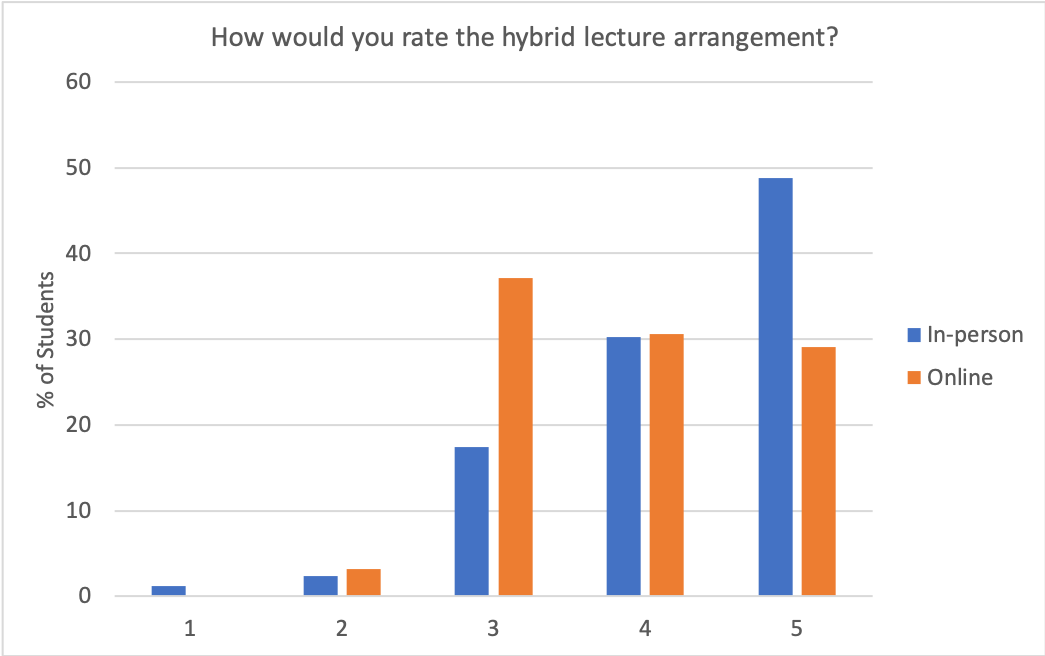}
   \caption{Participants' feedback on the hybrid arrangements.}
    \label{fig:hybrid}
  \end{center}
\end{figure}

\begin{figure}[!htbp]
 \begin{center}
  \includegraphics[width=\textwidth]{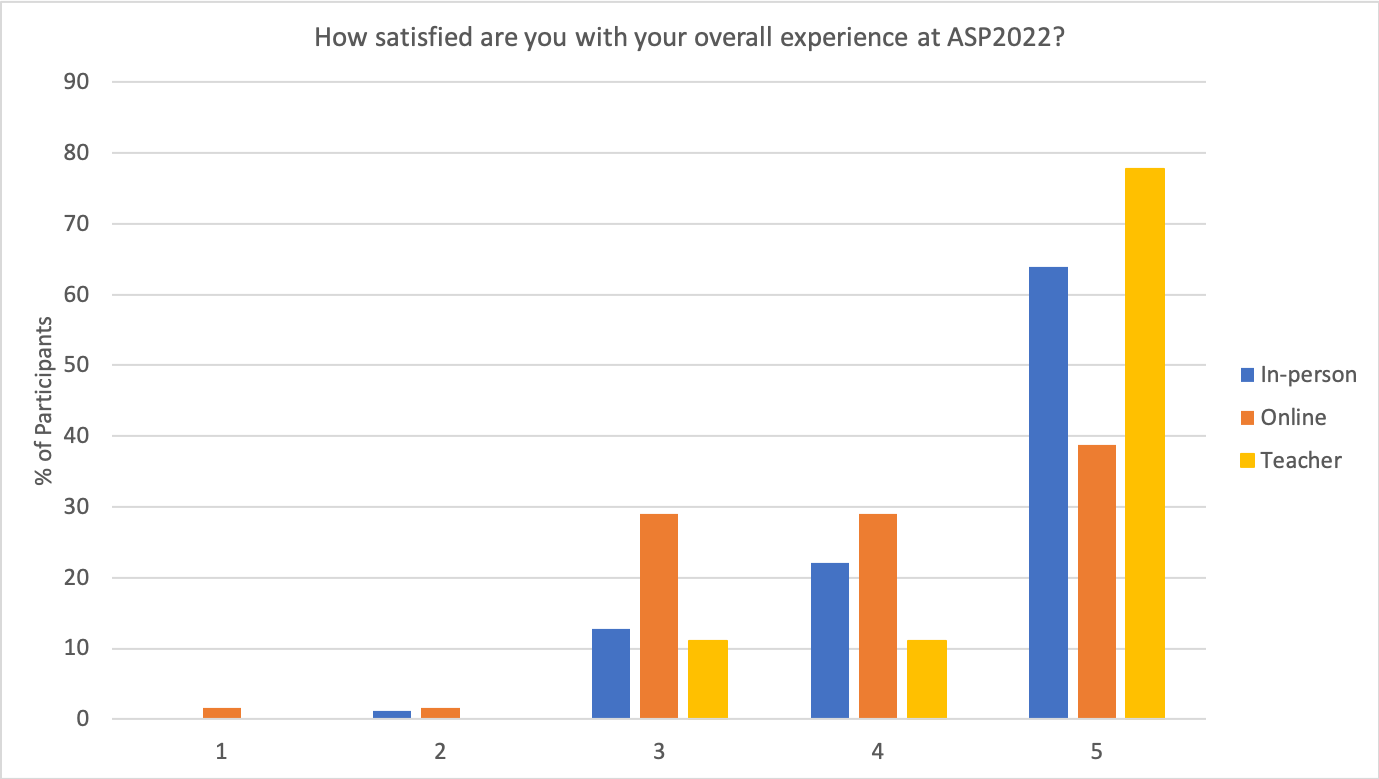}
   \caption{Participants were generally satisfied with the experiences at ASP2022.}
    \label{fig:satisfaction}
  \end{center}
\end{figure}

\begin{figure}[!htbp]
 \begin{center}
  \includegraphics[width=\textwidth]{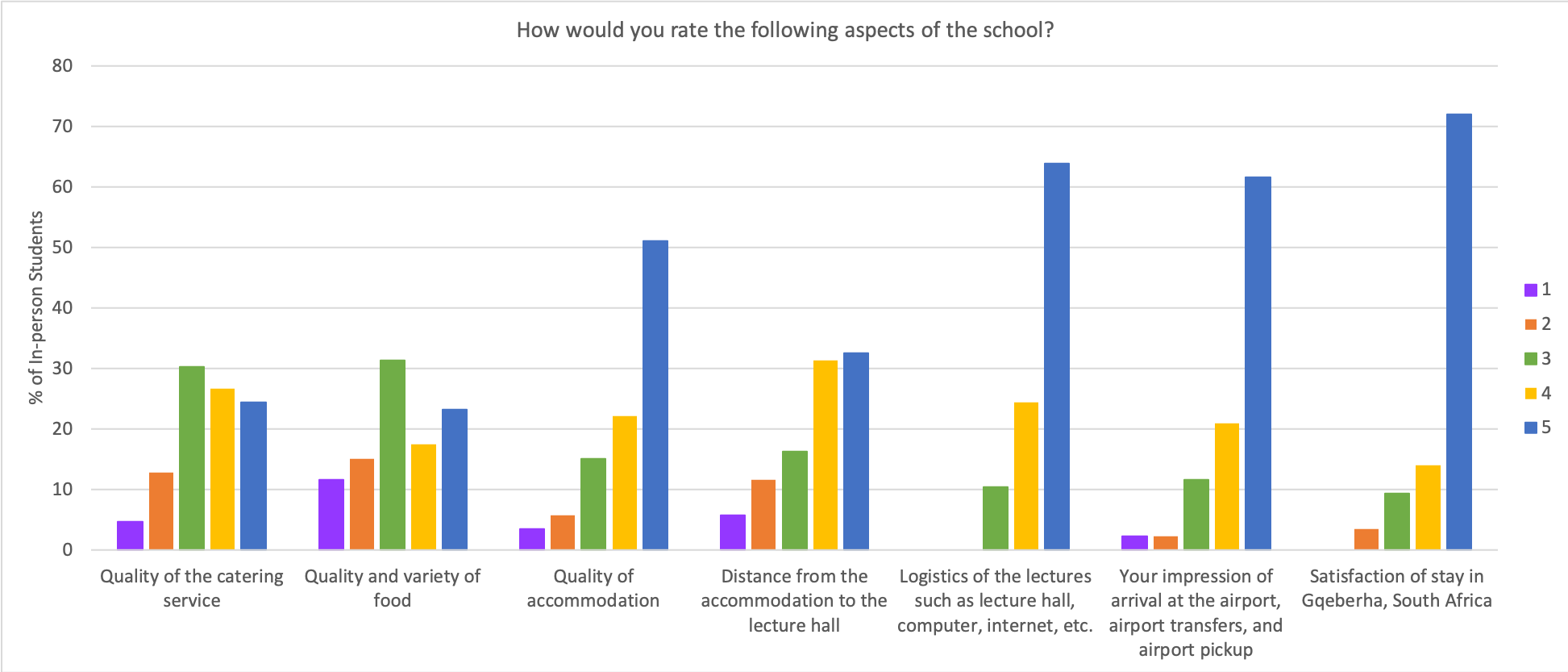}
   \caption{Feedback on various aspects of the logistics.}
    \label{fig:aspects}
  \end{center}
\end{figure}

\begin{figure}[!htbp]
 \begin{center}
  \includegraphics[width=\textwidth]{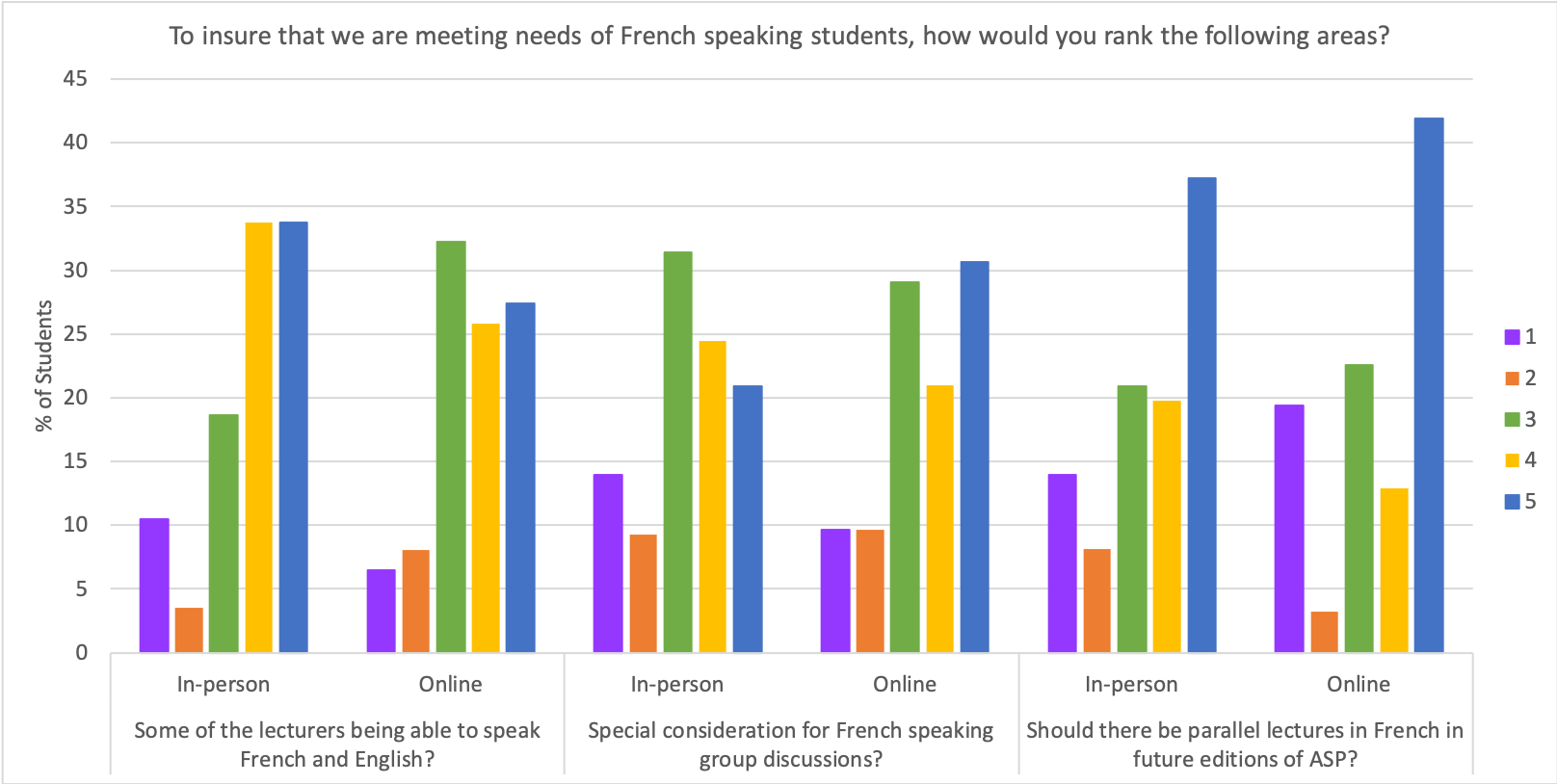}
   \caption{Feedback on the language of engagement. Consideration for French speaking participants was generally preferred.}
    \label{fig:language}
  \end{center}
\end{figure}

\begin{figure}[!htbp]
 \begin{center}
  \includegraphics[width=0.5\textwidth]{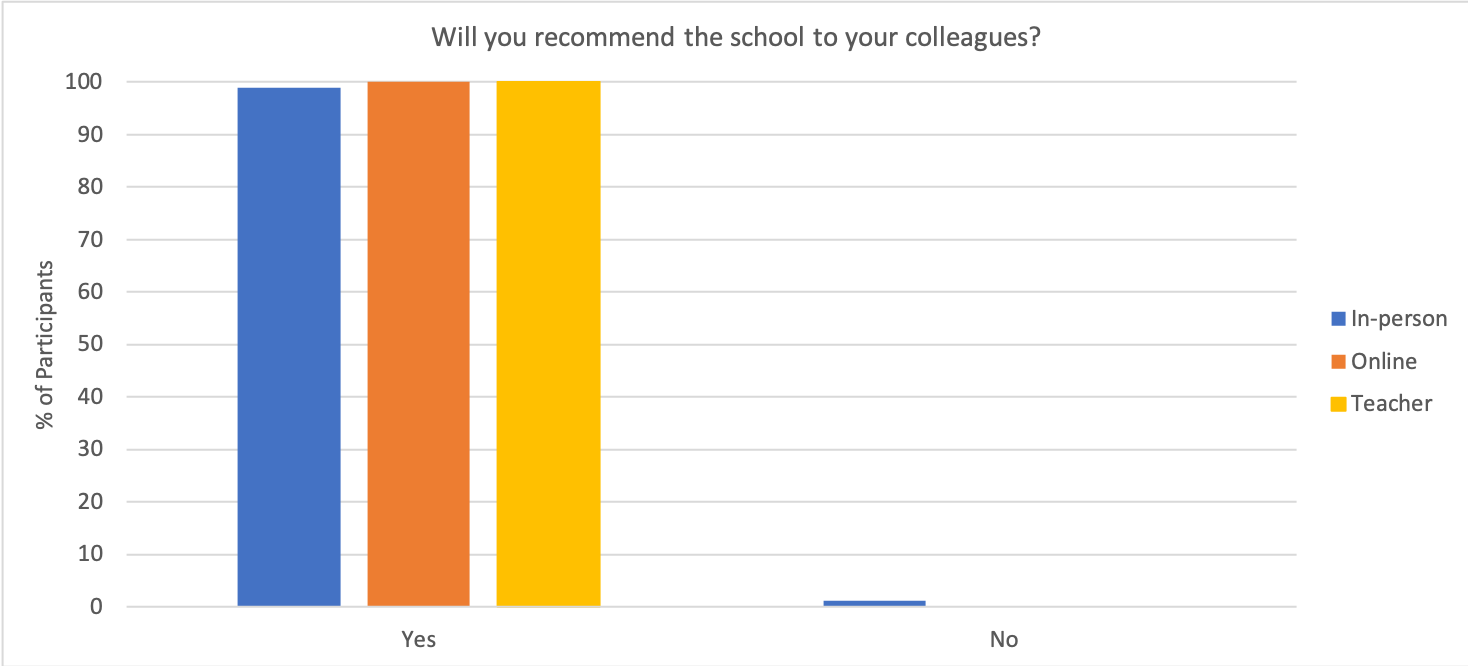}
  \includegraphics[width=0.45\textwidth]{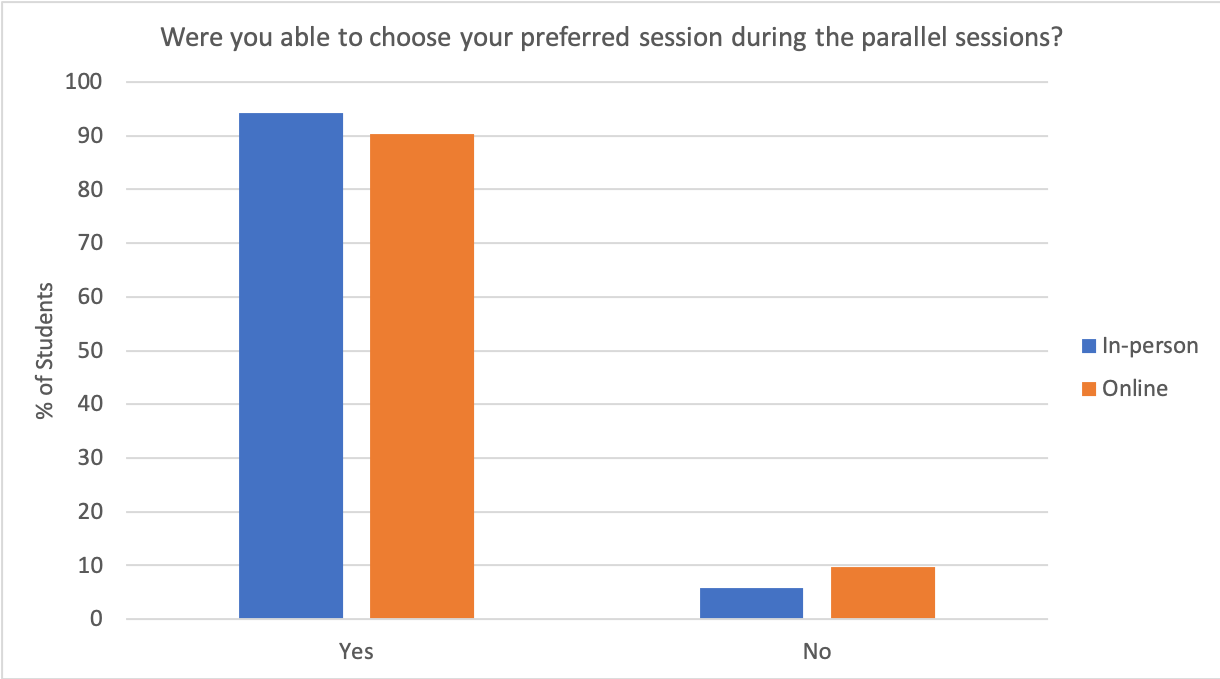}
  \includegraphics[width=0.45\textwidth]{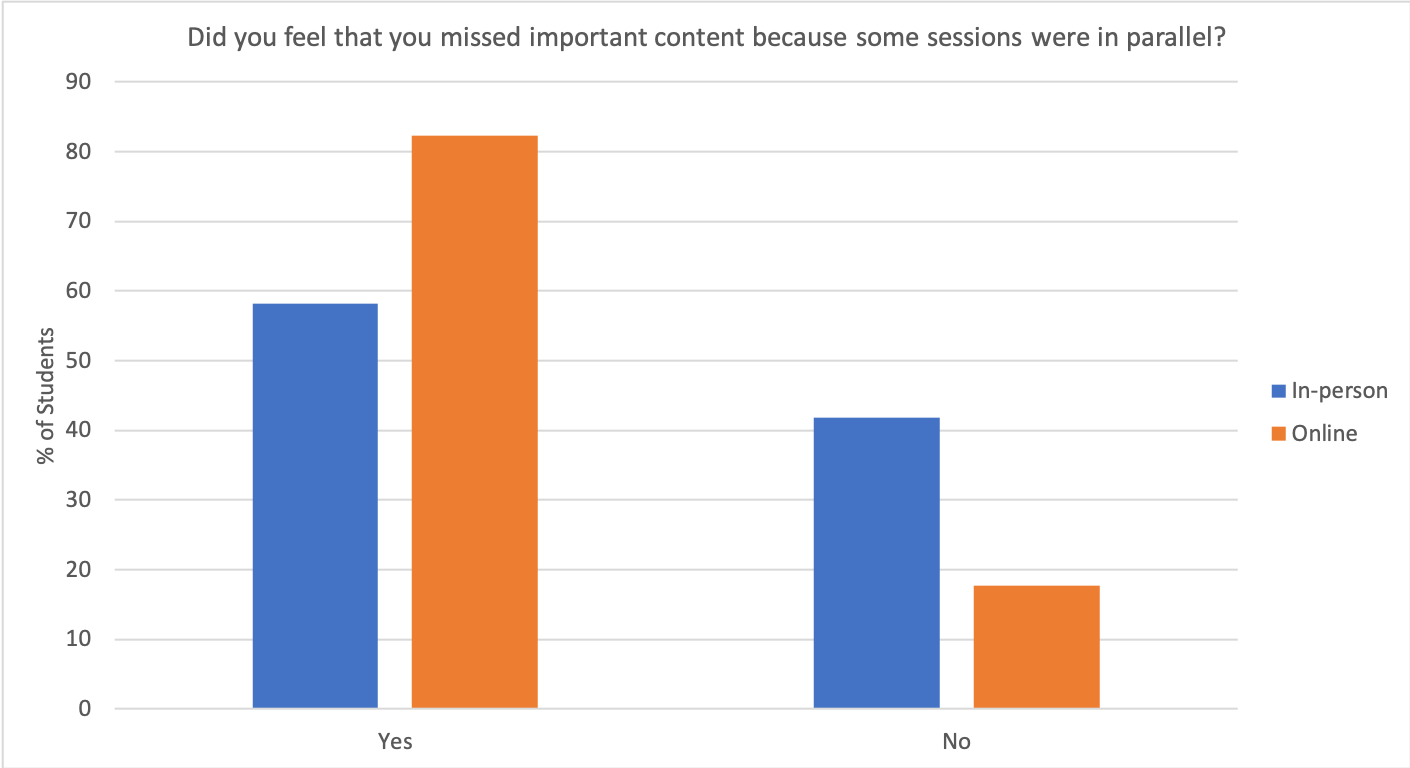}
  \includegraphics[width=0.45\textwidth]{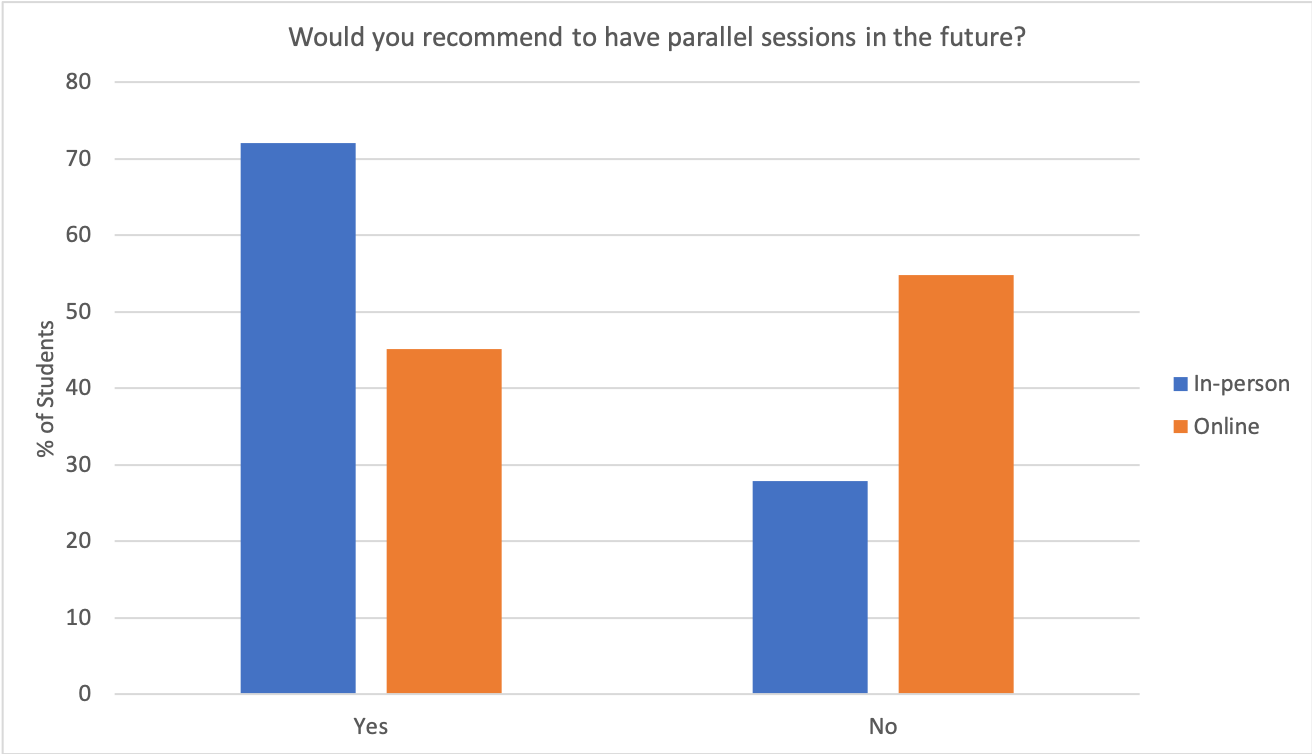}
   \caption{Most participants will recommend ASP to colleagues. However, parallel activities should be improved.}
    \label{fig:recommendation}
  \end{center}
\end{figure}

\begin{figure}[!htbp]
 \begin{center}
  \includegraphics[width=\textwidth]{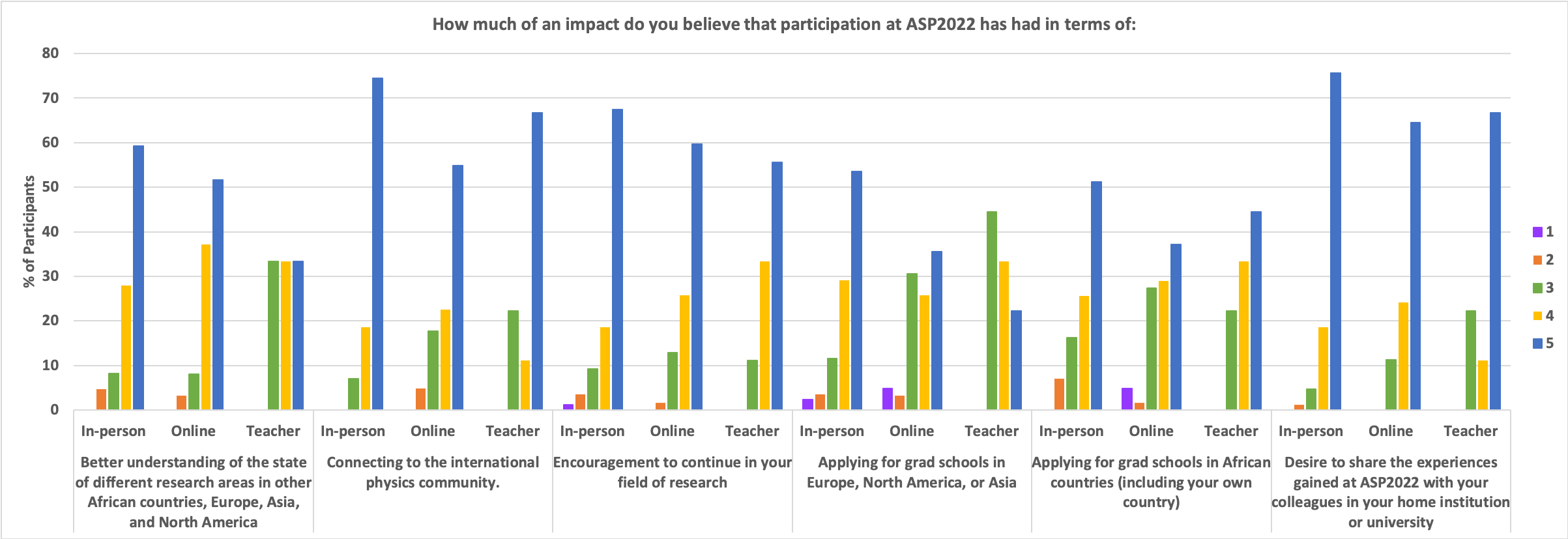}
   \caption{ASP2022 had positive impacts in various aspects of academic engagement.}
    \label{fig:impacts}
  \end{center}
\end{figure}

\begin{figure}[!htbp]
 \begin{center}
  \includegraphics[width=\textwidth]{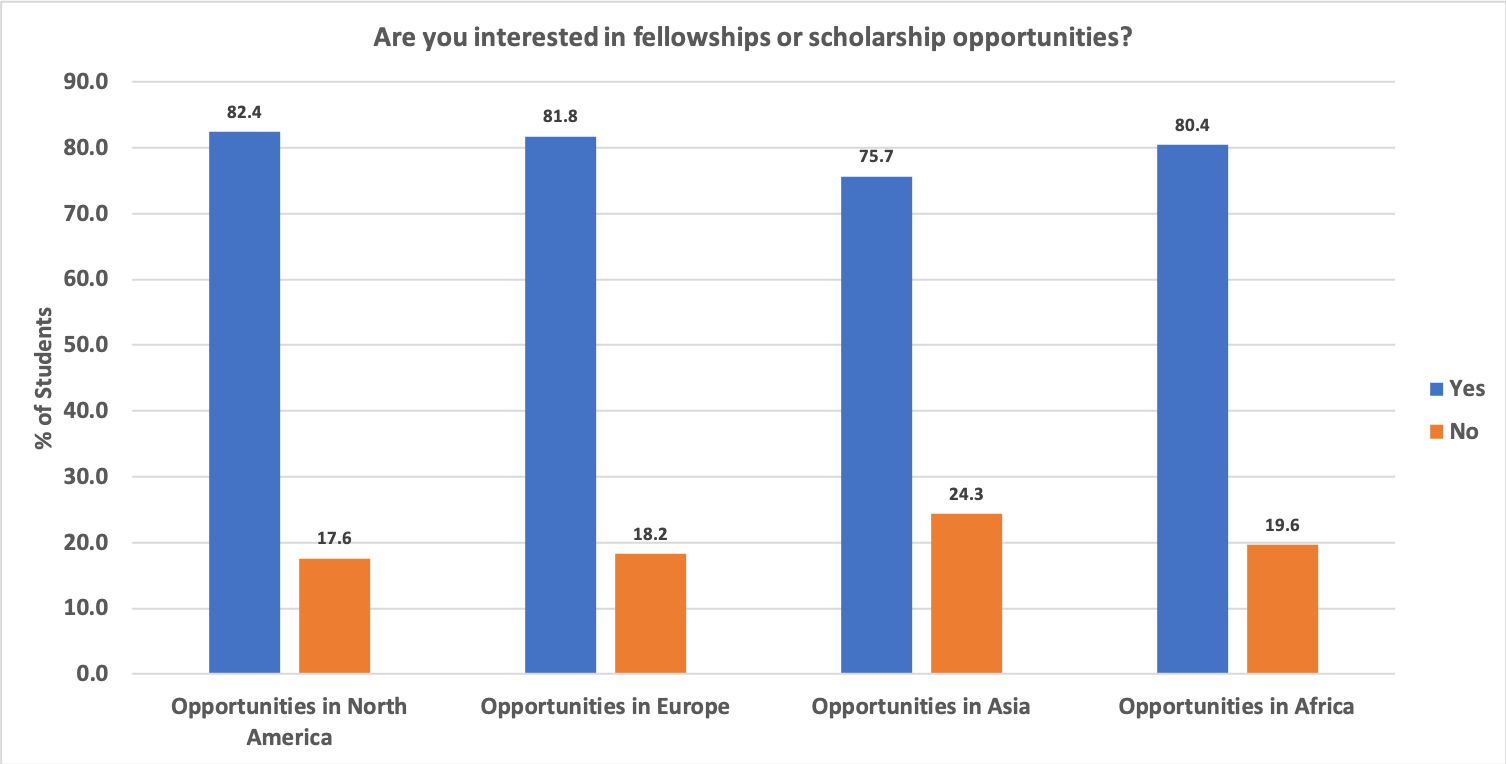}
   \caption{Participants expressed interest in higher education opportunities.}
    \label{fig:fellowships}
  \end{center}
\end{figure}

\begin{figure}[!htbp]
 \begin{center}
  \includegraphics[width=\textwidth]{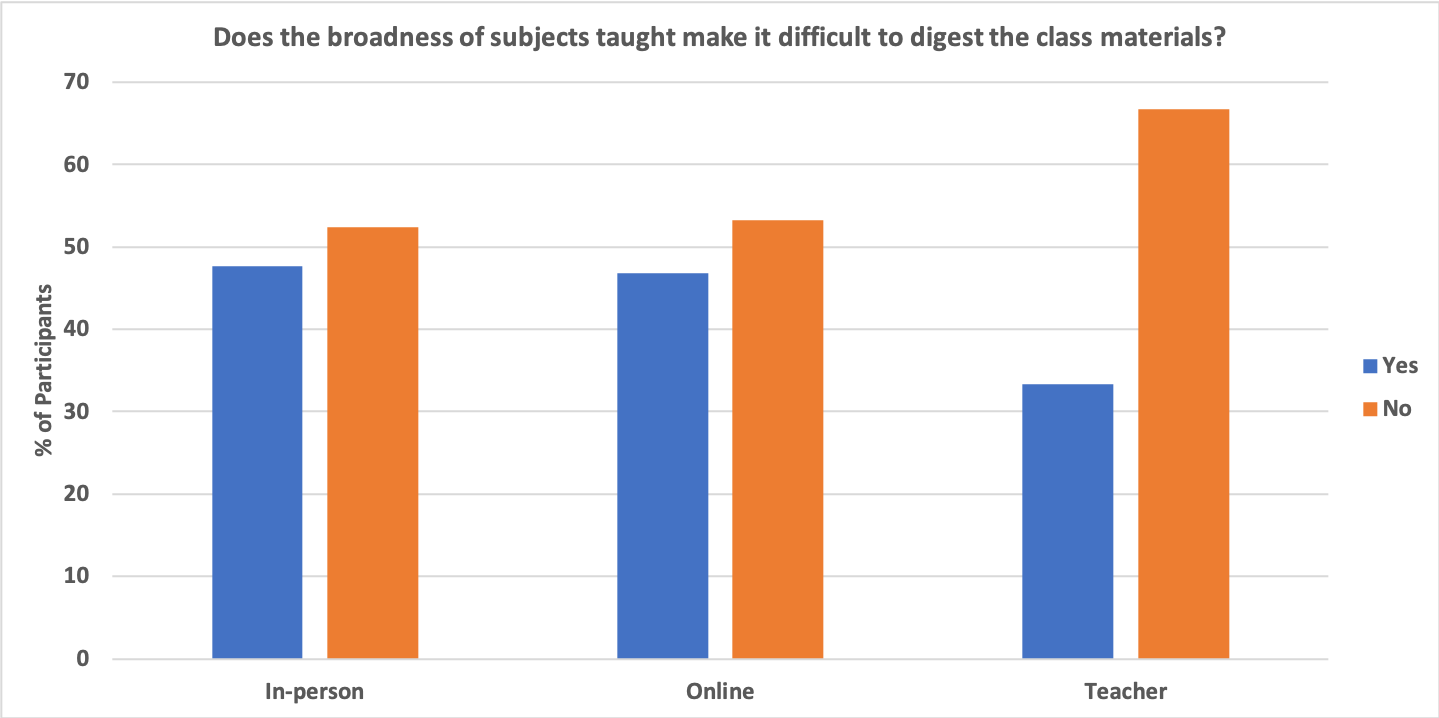}
   \caption{The broadness of the subjects taught posed difficulties for some participants.}
    \label{fig:digest}
  \end{center}
\end{figure}

\begin{figure}[!htbp]
 \begin{center}
  \includegraphics[width=\textwidth]{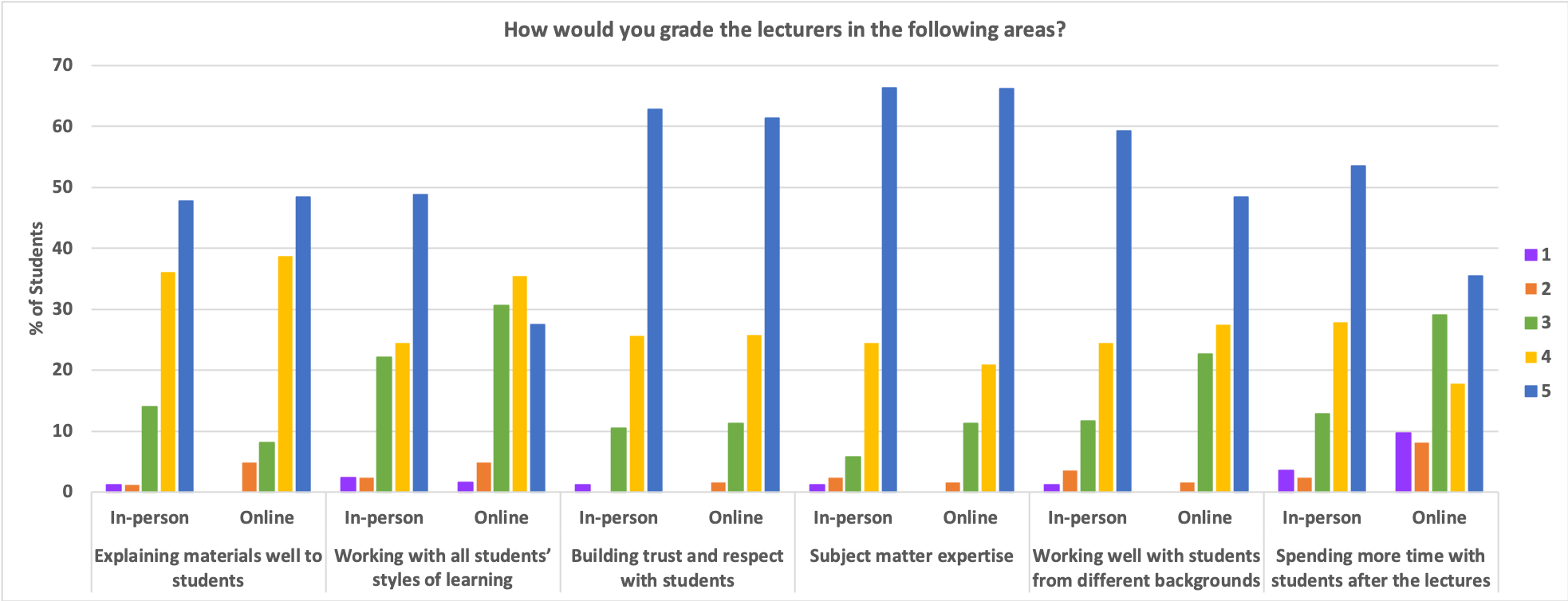}
   \caption{Participants reported positive interactions with lecturers.}
    \label{fig:lecturers}
  \end{center}
\end{figure}

\begin{figure}[!htbp]
 \begin{center}
  \includegraphics[width=\textwidth]{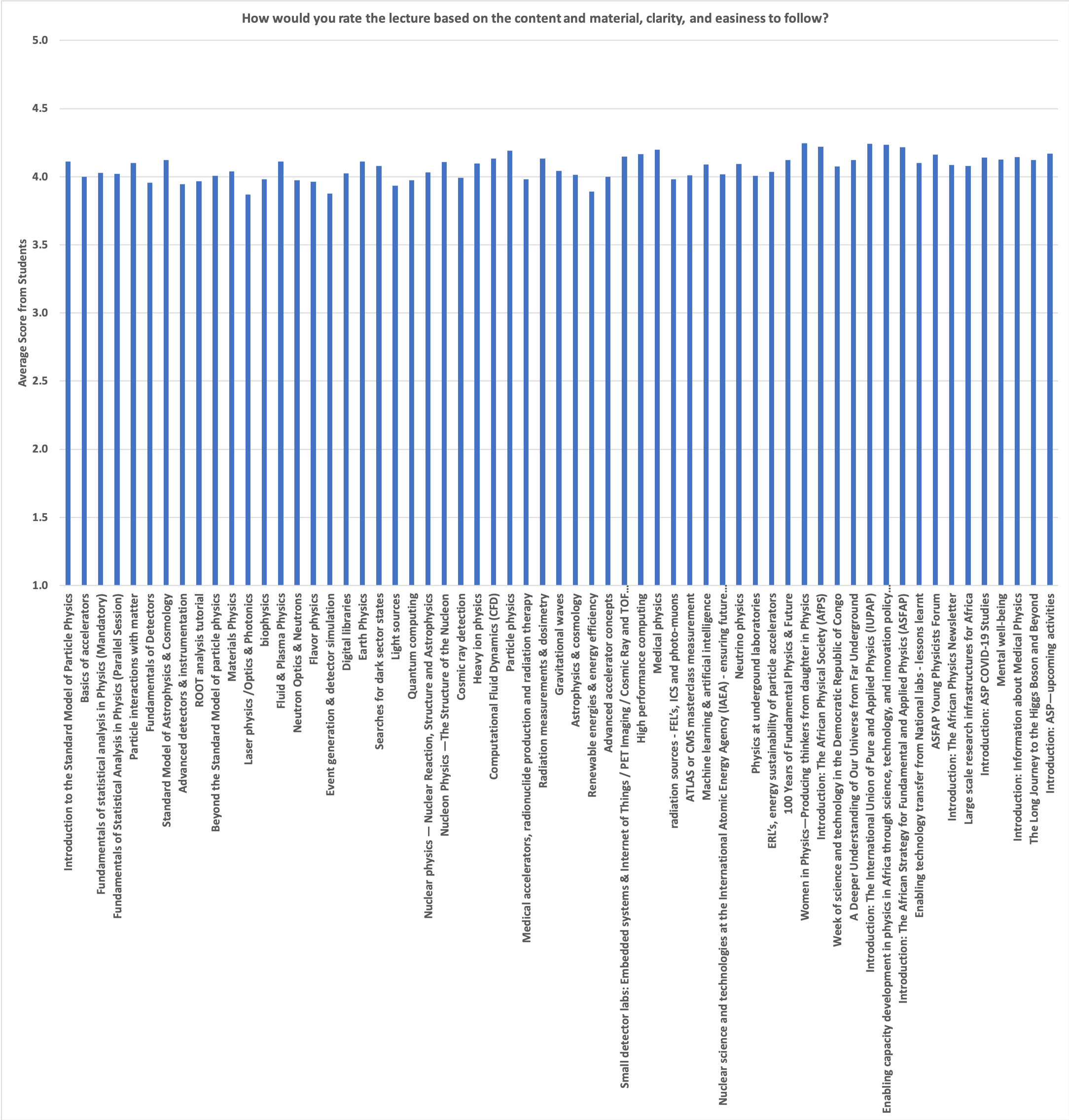}
   \caption{Participants rated favorably the lectures that they attended.}
    \label{fig:clarity}
  \end{center}
\end{figure}

Towards the end of the event, we asked participants for feedback through a survey. One hundred and fifty-seven students and teachers responded to the survey; of these, sixty-two participated online. Figure~\ref{fig:hybrid} shows that participants appreciated the hybrid arrangement, although online participants were less satisfied; understandably, most online participants expressed desire for in-person participation.
As shown in Figure~\ref{fig:satisfaction}, most of participants were satisfied with their ASP2022 experiences.
In the details, participants were satisfied with various aspects of logistics, as shown in Figure~\ref{fig:aspects}. However, the feedback shows that the quality of the catering services and diversity of meals could have been better.
In feedback presented in Figure~\ref{fig:language}, diversity in the languages of instructions and engagement should be entertained in future ASP considering the backgrounds of the participants.
Most survey respondents said they will recommend ASP to colleagues; however, improvements are needed in the organization of the parallel activities, as shown in Figure~\ref{fig:recommendation}.
In Figure~\ref{fig:impacts}, participants reported that ASP2022 had positive impacts on various aspects of their academic efforts.
Most participants were interested in fellowship opportunities for higher education, as shown in Figure~\ref{fig:fellowships}. 
Many participants reported that the broadness of the subjects presented made it difficult to digest lecture materials, as shown in Figure~\ref{fig:digest}; this issue is further discussed in Section~\ref{sec:out}.
In Figure~\ref{fig:lecturers}, participants were satisfied with their interactions with lecturers.
For the lectures that they did follow, participants rated favorably the content, material, clarity and easiness to follow, as shown in Figure~\ref{fig:clarity}.
The articles of Ref.~\cite{ASP2022-articles} offer more feedback based on interviews and discussions during ASP2022.

\section{Outlook}
\label{sec:out}

Going forward, improvements will be implemented in a number of key areas to make the ASP experience more enriching and successful. Resources and logistics do not allow for in-person participation of all the selected candidates. Organizing ASP in hybrid format seems to be a way to increase participation. More resources, both in staff and equipment will be allocated to make the hybrid format effective.  

Considering the academic concentrations of participants as shown in Figure~\ref{fig:majors}, we designed the program in plenary lectures, followed by more topical and advanced lectures in parallel sessions. Participants were expected to select the lectures that best supported their academic needs; the selections were given to the lecturers before the event.  Still, the parallel lectures drew a mixture of students with various levels of training in the subjects. Lecturers will have to consider the backgrounds of their audience, as shown in figures~\ref{fig:degrees} and~\ref{fig:majors}, in the design and delivery of their materials, even the more advanced parallel interactions. 

ASP is normally a three-week event organized as a two-week school, supplemented with the conference in the third week. This arrangement often led to confusion whether the event was a school or a conference. In the future, the school  and the conference parts of ASP will be permanently split and organized in different countries in alternating years—such a split will also allow more African countries the opportunity to host an ASP event.

\section{Conclusions}
\label{sec:conc}
ASP2022 was organized on November 28 to December 9, 2022 at Nelson Mandela University in Gqeberha, South Africa. It was as a hybrid event. In terms of the number of participants, ASP2022 was the largest ASP event to-date: 231 high school students, 76 high school teachers, 191 students---supported by many lecturers and organizers---attended the event. The duration of ASP2022 was shortened to two weeks compared to three in the previous years; however, the scientific content did not diminish. The scientific program was arranged in plenary and parallel sessions for students of various majors and teachers to select sets of courses that best supported their academic growth. The scientific content was developed in fundamental and applied physics through lectures, hands-on experimentation and computing tutorials. The program also included a forum for discussion with policymakers on the sustainability of ASP, capacity development and retention. 

The first ASP, in a hybrid environment, presented challenges in the running of the running of event; however, feedback from participants was generally positive. Through a survey, we noted all the aspects to be improved in future events. At the time of writing, the next upcoming ASP event will  be the third African Conference on Fundamental and Applied Physics, ACP2023, planned at Nelson Mandela University, George Campus, on September 25--29, 2023~\cite{acp2023}: registration and abstract submission are open.

\section*{Acknowledgments}
We thank all the sponsors of ASP2022 and the institutes who supported travels for lecturers. We also thank the organizing committee (local and international), in particular Brian Masara (SAIP), Dolly Ntintili (NMU), Reatile Mosia (NMU), Sisipho Ngesi (NMU), Jade Alexander (NMU), Andre Venter (NMU), Nobom Hashe (NMU)  and Hsien-Chang Albert Liu (NMU). We appreciate the efforts of Susanne Henningsen (ICTP) with the management of student application data and arrangement of international travels. We are grateful for the tremendous efforts of the lecturers. We acknowledge the collegial atmosphere maintained by all the participants, i.e. students, teachers, lecturers, pupils, organizers and policymakers.

\newpage

\bibliographystyle{elsarticle-num}
\bibliography{myreferences} 

\end{document}